\documentclass[aps,prd,tightenlines,preprint,groupedaddress,showpacs,showkeys]{revtex4}

\usepackage{graphicx}

\begin{document}

\preprint{KANAZAWA-04-01}

\title{~\\~\\~\\~\\Infrared Non-perturbative Propagators of Gluon and Ghost via Exact Renormalization Group\\~\\~\\
}


\author{Junya Kato}
\email[email: ]{jkato@hep.s.kanazawa-u.ac.jp}
%\thanks{}
\affiliation{Institute for Theoretical Physics,
Kanazawa University Kanazawa 920-1192, Japan}


\date{\today}

\begin{abstract}
The recent investigations of pure Landau gauge SU(3) Yang-Mills theories 
which are 
based on the truncated Schwinger-Dyson equations (SDE) indicate an infrared 
power law behavior of the gluon and the ghost propagators. It has been shown 
that the gluon propagator vanishes (or finite) in the infrared limit, while 
the ghost propagator is more singular than a massless pole, 
and also that there 
exists an infrared fixed point of the running gauge coupling. 
%These infrared power solutions are characterized by one critical exponent 
% (ghost anomalous dimension) $\kappa$ and the infrared fixed point of running 
%coupling% $\alpha^*$. 
In this paper we reexamine this picture by means of the exact 
(non-perturbative) renormalization group (ERG) equations under some
approximation scheme, in which we treat not only two point functions
but also four point vertices in the effective average action with 
retaining their momentum dependence.
Then it is shown that the gluon and the ghost propagators with the infrared
power law behavior are obtained as an attractive solution starting from 
rather arbitrary ultraviolet bare actions. 
Here it is found to be crucial to include the momentum dependent 
four point vertices in the ERG framework, since otherwise the RG flows 
diverge at finite scales.
The features of the ERG analyses in comparison with the SDE are also 
discussed.
\end{abstract}


\maketitle

\section{Introduction\label{introduction}}
In quantum field theories, the Green functions are the most fundamental 
objects in order to understand the dynamics. 
Especially in QCD, the two point functions tell us various informations 
about non-perturbative phenomena, such as dynamical mass generation of 
chiral fermions and also color confinement, which are expected to occur 
in the infrared region.
There have been many studies concerning to the dynamical chiral symmetry 
breaking (D$\chi$SB) in QCD and its effective models using SDE and ERG.
Though these are successful in describing D$\chi$SB, 
much less has been well known as for color confinement due to complicated
gluon self interactions .

The recent SDE studies for the gluon and the ghost propagators in Landau 
gauge Yang-Mills theories\cite{smekal_1, smekal_2, fischer_1, fischer_2, 
atkinson_1, bloch_1, kondo_1, kondo_2, zwanziger_1} initiated by Smekal, 
Hauck and Alkofer\cite{smekal_1}, indicate the power law behavior: the 
gluon propagator vanishes (or finite) in the infrared limit, while the 
ghost propagator is more singular than a massless pole. 
IR asymptotic forms of the propagators are found to be described by one 
exponent (ghost anomalous dimension) $\kappa$ as
\begin{equation}
D_Z(p^2)\sim\left(p^2\right)^{-1+2\kappa},
\quad \quad
D_G(p^2)\sim\left(p^2\right)^{-1-\kappa},
\label{1}
\end{equation}
for the gluon and the ghost propagators respectively.
The actual value of $\kappa$ obtained by solving the SDE is dependent on 
the approximation scheme. However fit of the data by lattice simulations 
at some finite Euclidean momentum region\cite{bloch_1} as well as axiomatic 
consideration \cite{kondo_1} indicates $\kappa\approx 0.5$.
It is also interesting that the solutions of the SDE show consistent 
behavior with the lattice data.
Implication of this behavior to confinement is attributed to
realization of the so-called Kugo-Ojima confinement scenario via the 
BRS quartet mechanism of colored particles \cite{kugo_ojima, kugo}. 
In the Landau gauge, the confinement criterion is understood as infrared
enhancement of the ghost propagator, and has been proven to be
fulfilled under general ansatz by using the SDE for ghost in the Landau 
gauge \cite{watson}.
In addition to the confinement criterion, the IR asymptotic power solutions 
satisfy  also the Gribov-Zwanziger horizon condition \cite{gribov, 
zwanziger_2}. 
The horizon condition is a dynamical consequence obtained in restricting 
the support of the Faddeev-Popov measure to the interior of the Gribov region 
to avoid gauge copies. The conditions are represented in terms of the 
propagators as 
\begin{equation}
\lim_{p^2\rightarrow 0}\,D_Z(p^2)=0,
\quad \quad
\lim_{p^2\rightarrow 0}\,
\left[p^2 D_G(p^2)\right]^{-1}=0,
\label{2}
\end{equation} 
which are found to be fulfilled by the SDE solutions.
Thus, one may say that the power solutions of the SDE most probably 
capture an aspect of confinement \cite{langfeld}, though the string 
tension has not been derived directly from the two point functions.

Here it would be worthwhile to mention also Zwanziger's observations 
that the cutoff prescription at the Gribov horizon resolves
an ambiguity in the solutions of SDE
\cite{zwanziger_1, smekal_2}. 
First it is noted that cutting off the functional integral at the Gribov
horizon does not alter the SDE.
Also the boundary contribution does not affect the form of SDE, because 
the Faddeev-Popov measure vanishes on the boundary. 
However, it is shown that the cutoff at the Gribov horizon provides 
supplementary conditions for the solutions of SDE so as to make the 
Faddeev-Popov measure positive definite. 
Indeed it is needed to know the infrared asymptotic power law behavior
of the solutions in the practical analyses of the SDE.
However there are found multiple solutions satisfying the power
law ansatz.
Then Zwanziger pointed out that infrared asymptotic solution in the
SDE analyses done so far has been implicitly chosen to satisfy the 
supplementary condition\cite{smekal_1, smekal_2,atkinson_1}.
%And also the horizon condition are used as renormalization condition 
%at $p^2=0$ on ghost SDE \cite{zwanziger_1, smekal_2}.
Thus, in the practical SDE analyses, it is important to impose proper
infrared boundary conditions in order to find out the physical solution
\footnote{It seems through numerical analysis that other solutions 
but the physical one cannot be connected smoothly to the perturbative
solution at the high momentum region. 
}.
Contrary to this, however, the exact renormalization group (ERG) approach 
does not need to use the supplementary conditions for the infrared 
behaviors, the power law solutions with a certain exponent emerge uniquely 
as will be shown in this paper. 

The ERG describes continuous evolution of the coarse grained effective 
action in the spirit of Wilson's renormalization group\cite{wilson}. 
In continuum field theories, this evolution is most conveniently expressed
in terms of the effective average action $\Gamma_\Lambda$ with the coarse 
grained (infrared cutoff) scale $\Lambda$ \cite{wetterich_2,wetterich_3}. 
In practice the ERG has been
frequently used in various non-perturbative calculations and 
found to be rather useful.
In contrast to the SDE framework, in the ERG, solving the flow equation 
starting with a ultraviolet bare action $\Gamma_{\Lambda_0}$ gives
a unique infrared solution $\Gamma_{\Lambda=0}$ as $\Lambda_0 \rightarrow 0$. 
Indeed it can be shown schematically that the solutions of ERG
satisfy the corresponding SDE on the exact generating functional 
level\cite {ellwanger_1, terao_1}. 
However the correspondence becomes totally unclear, once some approximation 
or truncation is performed, except for some rare cases\cite{aoki_1}.
Especially in the cases that the SDE allows multiple solutions, it would be 
an interesting problem to see whether the physical solution can be 
obtained by solving the ERG equations.
Another good feature of the ERG scheme is the infrared attractive
property of the RG flows.
In renormalizable quantum field theories, the same infrared effective
action is derived from rather arbitrary ultraviolet bare actions owing 
to the universality argument \cite{polchinski}.
Thus it is expected to be rather straightforward to analyze the infrared
two point functions of the Yang-Mills theories in the ERG framework.

In this paper, we investigate the infrared behavior of the effective action,
specially the gluon and the ghost propagators of Landau gauge 
Yang-Mills theory by solving the approximated ERG equation. 
The main purpose of this paper is to see, first whether the infrared power
behavior can emerge also in the ERG framework at all, and then which
kind of corrections are necessary to be taken into account for that.
Previously, Ellwanger, Hirsch and Weber also have studied infrared 
behavior of the propagators by using ERG \cite{ellwanger_1}, and they 
concluded that the confining $1/p^4$ behavior for the heavy quark potential
which was derived from the effective four quark interactions. 
However, their RG flows become singular at a finite scale, and therefore 
the infrared power law behavior has not been seen.  
The most different point with our analysis is inclusion of the four
point vertices among gluons and ghosts, which are necessary ingredient
to produce the power law solutions.
In their analysis, only the four gluon vertex, which was determined by using 
the Slavnov-Taylor identities, was included. 
Our analysis indicates that the effective four point vertices generated by box
diagrams play a specially important role, because the large ghost anomalous 
dimension enhances these four point vertices. 
Indeed, the infrared solutions with power behavior are found to be 
obtained by taking these into account, 
although with a small exponent $\kappa \approx 0.146$.

Application of the ERG to Yang-Mills theories immediately faces problems 
of the gauge  or the BRS invariance, because the infrared cutoff of momentum
breaks the local symmetry. 
Schematically, this problem may be managed by use of the modified 
Slavnov-Taylor identities (mSTi) 
\cite{ellwanger_brs,ellwanger_1,ellwanger_2},
which guarantee the broken BRS symmetry
of the effective average action $\Gamma_\Lambda$ at an intermediate infrared
cutoff scale $\Lambda \ne 0$ to recover the symmetry at infrared limit $\Lambda \rightarrow 0$. 
There the BRS non-invariant counter terms, such as gluon mass, should be
controlled by the mSTi. However, the infrared BRS invariance is not 
maintained any more once we truncate the effective action.
Effectiveness of the mSTi in the practical sense is unclear apart
from perturbative analyses.
Therefore, in this paper we do not pursuit for the problems of
gauge invariance and simply discard the BRS non-invariant corrections. 
This is a difficult aspect of non-perturbative analysis of gauge theories
not only in the ERG but also in the SDE framework.

Furthermore, for the sake of simplicity, we neglect also corrections
with the three gluon vertex in the flow equations, 
though the resultant flows will no longer be 
Yang-Mills theory's ones.
However, according to the previous SDE 
analyses \cite{smekal_1, smekal_2, fischer_1, fischer_2, zwanziger_1}, 
contribution of the three gluon vertex was found to be sub-leading 
as far as  the IR asymptotic solutions are concerned. 
Therefore we also adopt such a simple approximation as the first trial
to apply the ERG equation for the infrared dynamics of Yang-Mills
theories.

The paper is organized as follows: 
In section 2, we explain briefly the exact flow equation for Landau gauge
Yang-Mills theory and also introduce some basic notations. 
In section 3, we present our approximation scheme and give the practical
RG equations, which we shall analyze in the followings in detail. 
Results of the numerical analysis are presented in section 4.
There the RG flows of the momentum dependent gluon and ghost form factors 
and also the running gauge coupling obtained in our scheme are shown. 
Section 5. is devoted for some discussions and remarks on further issues. 
Some comparisons between the ERG and the SDE formalisms are also 
summarized there.

\section{Exact Renormalization Group\label{exactrenormalizationgroup}}
The exact renormalization group is a realization of the Wilson
renormalization group transformation in continuum field theories, 
and there have been known various formulations. 
The most commonly used form is the so-called flow equation written 
in terms of the effective average action $\Gamma_\Lambda$\cite{wetterich_2}.
The effective average action $\Gamma_\Lambda$ is the effective action 
$\Gamma$ obtained after only quantum fluctuations with momentum scale 
$p^2 \ge \Lambda^2$ are integrated out. Therefore, it is an efficient 
tool interpolating the classical bare action $S$ and the quantum effective 
action $\Gamma$;
\begin{equation}
\lim_{\Lambda \rightarrow \Lambda_0}\Gamma [\mathbf{\Phi};\Lambda] = 
S_{\rm bare}[\mathbf{\Phi}] \quad,\quad\lim_{\Lambda\rightarrow 0}
\Gamma[\mathbf{\Phi};\Lambda] = \Gamma[\mathbf{\Phi}]\label{3}.
\end{equation}
Here dependence on the ultraviolet cutoff $\Lambda_0$ is 
suppressed implicitly. 
Explicitly the effective average action $\Gamma_\Lambda$ is defined through 
the Legendre transformation of the IR regularized generating functional for 
the connected Green functions $W_\Lambda$;
\begin{equation}
W [\mathbf{J};\Lambda]=\log \int{\cal D}\mathbf{\Phi}
\exp\left\{-S[\mathbf{\Phi}]-\Delta S[\mathbf{\Phi};\Lambda]+
\mathbf{J}\cdot\mathbf{\Phi}  \right\},
\label{4}
\end{equation}
where $\Delta S[\mathbf{\Phi};\Lambda]$ is the IR cutoff term introduced 
to suppress the infrared mode of $p^2<\Lambda^2$, and which is normally 
quadratic in fields $\mathbf{\Phi}$. 
The Legendre transformation of $W_\Lambda$ is defined by  
\begin{equation}
\Gamma[\mathbf{\Phi};\Lambda] = 
-W[\mathbf{J};\Lambda]+\mathbf{J}\cdot\mathbf{\Phi}-
\Delta S[\mathbf{\Phi};\Lambda].
\label{5}
\end{equation}
The flow equation is given by infinitesimal variation of 
$\Gamma_\Lambda$ with respect to  the IR cutoff scale $\Lambda$ 
and it can be written down as one-loop exact form. 
In the case of pure Yang-Mills theory, the flow equation may be given as
\begin{equation}
\partial_\Lambda \Gamma [\mathbf{\Phi};\Lambda] = 
\frac{1}{2} {\bf Str} \left[\partial \mathbf{R} \cdot 
\left(\frac{\delta^2 \Gamma}{\delta \bar{\mathbf{\Phi}} 
\delta \mathbf{\Phi}} [\mathbf{\Phi} ;\Lambda]  
+\mathbf{R} \right)^{-1}\right],
\label{6}
\end{equation}
where we introduced the matrix notation following Ellwanger 
{\it et al.}\cite{ellwanger_2}.
The supertrace includes momentum integration as well as summation
of color and Lorentz indices. 
The field $\mathbf{\Phi}$ and $\bar{\mathbf{\Phi}}$ denote
the gluon, ghost and anti-ghost fields in short hand notation as 
\begin{equation}
\bar{\mathbf{\Phi}} = \left(A,c,\bar{c}\right), \quad 
\mathbf{\Phi} = \left(A,-\bar{c},c\right).
\label{7}
\end{equation}
The symbol $\mathbf{R}$ stands for the diagonal matrix of the 
cutoff functions, which are explicitly given by
\begin{eqnarray}
\mathbf{R}(p) &=& 
{\rm diag}\left(R^{ab}_{\mu\nu}(p;\Lambda)\,,\,
-\tilde{R}^{ab}(p;\Lambda)\,,\,
-\tilde{R}^{ab}(p;\Lambda)\right), \\
\label{8}
\partial \mathbf{R}(p) &=& 
{\rm diag}\left(\partial_\Lambda R^{ab}_{\mu\nu}(p;\Lambda)\,,\,
-\partial_\Lambda \tilde{R}^{ab}(p;\Lambda)\,,\,
-\partial_\Lambda \tilde{R}^{ab}(p;\Lambda)\right).
\label{9}
\end{eqnarray}
The cutoff term added in the definition of the cutoff
generating functional given by eq.~(\ref{4}) is written down
in terms of these functions as
\begin{equation}
\Delta S [\mathbf{\Phi} ;\Lambda] = 
\frac{1}{2}\int_p A^a_\mu (-p) R^{ab}_{\mu\nu}(p;\Lambda)A^b_\nu (p)\,
-\,\int_p \bar{c}^a(-p)\tilde{R}^{ab}(p;\Lambda) c^b(p).
\label{10}
\end{equation}
Here we adopt the cutoff functions of the following form; 
\begin{eqnarray}
R_{\mu\nu}^{ab}(p;\Lambda) &=& 
\left[\left(\delta_{\mu\nu}-\frac{p_\mu p_\nu}{p^2}\right)R_\Lambda(p^2)+
\frac{1}{\xi(\Lambda)}R_\Lambda(p^2)\frac{p_\mu p_\nu}{p^2}\right]\,
p^2\delta^{ab},
\label{11}\\
\tilde{R}^{ab}(p;\Lambda) &=& 
R_\Lambda(p^2)p^2\delta^{ab},
\label{12}
\end{eqnarray}
for the gluon and the ghost fields respectively. 
We may choose a general cutoff functions for $R_\Lambda$ here, 
however in the following analyses, we will take sharp cutoff
limit given by,
\begin{equation}
R_\Lambda(p^2) = 
\lim_{\alpha\rightarrow\infty} \left(\frac{p^2}{\Lambda^2}\right)^{-\alpha},
\label{13}
\end{equation}
for the calculational simplicity.

\section{Approximation Scheme\label{approximationscheme}}

In the practical analysis, we need to perform some approximation 
in order to solve the flow equation schematically given by (\ref{6}). 
In the ERG framework it is done typically by truncating interactions in 
$\Gamma_\Lambda$ at some finite order, which should be compared with 
truncation of equations in the SDE formalism. 
We may improve the approximation systematically by 
increasing the interactions to be taken in.
The recent analyses of the SDE treat only the two point functions
of the gluon and ghost fields as dynamical ones. 
However it does not mean that we may truncate the effective action
or corrections to the action at the two point function order.
Although the solutions of the ERG and the SDE should coincide to each other 
at the full generating functional level, the truncation ruins the 
correspondence. 
As we will show in the subsequent section, it is necessary to include 
the four point vertices to realize the infrared finite power solution 
in the ERG formalism. 

One of our main interests is to see how the infrared power behavior of 
the gluon and ghost propagators emerge in the ERG framework. 
We shall now explain briefly the approximation scheme which we adopted
for this purpose. 
Our approximation scheme is, in short, to treat all diagrams that 
contribute to the gluon and ghost two point functions. 

Because the flow equation (\ref{6}) is one loop exact, the vertices 
contributing to the corrections of the two point functions are not more 
than four point.
Therefore, we truncate $\Gamma_\Lambda$ at the forth power of fields and
the explicit form is given as follows,  
\begin{eqnarray}
\Gamma\left[A,\bar{c},c;\Lambda\right] &=&
\frac{1}{2}\int_p\, \Gamma_{\mu\nu}^{ab}(p;\Lambda)\,
A_\mu^a (-p)\,A_\nu^b (p)\nonumber\\
&&+\int_{p,q,r}\!\!\!(2\pi)^4\delta^4(p+q+r)\,
V_{\mu\nu\rho}^{abc}(p,q,r;\Lambda)\,
A_\mu^a(p)A_\nu^b(q)A_\rho^c(r)\nonumber\\
&&+\int_{p,q,r,s}\!\!\!\!\!\!(2\pi)^4\delta^4(p+q+r+s)\,
V_{\mu\nu\rho\sigma}^{abcd}(p,q,r,s;\Lambda)\,
A_\mu^a(p) A_\nu^b(q) A_\rho^c(r) A_\sigma^d(s)\,\nonumber\\
&&-\int_p\,\tilde{\Gamma}^{ab}(p;\Lambda)\,
\bar{c}^a(-p) c^b(p)\nonumber\\
&&-\int_{p,q,r}\!\!\!(2\pi)^4\delta(p+q+r)\,
T_\mu^{abc}(p|q,r;\Lambda)\,
A_\mu^a(p) \bar{c}^b(q) c^c(r)\nonumber\\
&&+\int_{p,q,r,s}\!\!\!\!\!\!(2\pi)^4\delta^4(p+q+r+s)\,
T_{\mu\nu}^{abcd}(p,q|r,s;\Lambda)\,
A_\mu^a(p) A_\nu^b(q) \bar{c}^c(r) c^d(s)\nonumber\\
&&+\int_{p,q,r,s}\!\!\!\!\!\!(2\pi)^4\delta^4(p+q+r+s)\,
T^{abcd}(p,q,r,s;\Lambda)\,
\bar{c}^a(p) c^b(q) \bar{c}^c(r) c^d(s)\nonumber\\
&&+{\cal O}(\mathbf{\Phi}^5)\qquad
\left(\int_{p,\cdots, q}\equiv\int\!\!
\frac{d^4p}{(2\pi)^4}\cdots\int\!\!
\frac{d^4q}{(2\pi)^4}\right),
\label{14}
\end{eqnarray}
where the vertices are defined (anti)symmetric, {\it e.g}
$V_{\mu\nu\rho\sigma}^{abcd}(p,q,r,s;\Lambda)=
V_{\nu\mu\rho\sigma}^{bacd}(q,p,r,s;\Lambda)=\cdots$, and so on. 
Substituting the action (\ref{14}) to the flow equation (\ref{6}) 
and expanding the RHS of the flow equation in powers of fields, then
we obtain the coupled flow equations for the momentum dependent 
vertex functions.

However we may simplify the equations by using more specific form 
of the two point functions. 
For the gluon two point function, we choose,  
\begin{equation}
\Gamma_{\mu\nu}^{ab}(p;\Lambda) = 
\left[\left(\delta_{\mu\nu}-\frac{p_\mu p_\nu}{p^2}\right) 
f_Z(p^2;\Lambda)+\frac{1}{\xi(\Lambda)}\frac{p_\mu p_\nu}{p^2}\right]\,
p^2\delta^{ab}.
\label{15}
\end{equation}
Although we write also the $\Lambda$-dependent gauge parameter $\xi(\Lambda)$, 
we will consider only the Landau gauge limit case $\xi(\Lambda)\rightarrow0$ 
in this paper. It is noted that this limit can be taken consistently, 
because the Landau gauge is a fixed point $\xi^*(\Lambda)=0$
\cite{ellwanger_1, ellwanger_2} under the RG evolution. 

To be more precise, the equation (\ref{15}) is incomplete even in the 
Landau gauge, because the cutoff term (\ref{11}) induces the gluon mass 
term $\sim m(\Lambda)^2A_\mu^a A_\mu^a$ in the averaged effective action
due to lack of manifest BRS invariance.
Such BRS non-invariant terms must be controlled by the mSTi consistently
with the RG evolution\cite{ellwanger_1}. 
Here, however, we neglect simply such BRS non-invariant terms and also the 
corrections to them.
Then by contracting the flow equation for the gluon two point function 
with Brown-Pennington tensor 
$B_{\mu\nu}(p)=(\delta_{\mu\nu}-4p_\mu p_\nu /p^2)$, 
we may project out correction to the gluon mass term. 
For the ghost two point function, we choose, 
\begin{equation}
\tilde{\Gamma}^{ab}(p;\Lambda) = 
\delta^{ab}p^2 f_G(p^2;\Lambda).
\label{16}
\end{equation}
Thus the two point functions are reduced to $\Lambda$ dependent one 
component functions $f_Z$ and $f_G$. 
Integrating the flow equations toward 
$\Lambda\rightarrow 0$, the gluon and the ghost propagators, 
which are of our present interest, are found to be  
\begin{equation}
D_{Z,\mu\nu}(p^2)=
\frac{T_{\mu\nu}(p)}{p^2 f_Z(p^2;\Lambda\rightarrow0)}\quad,\quad 
D_G(p^2)= -\frac{1}{p^2f_G(p^2;\Lambda\rightarrow0)},
\label{17}
\end{equation}
where $T_{\mu\nu}(p)=(\delta_{\mu\nu}-p_\mu p_\nu /p^2)$ is the 
transverse projection. 

Next we restrict the three gluon vertex and the three ghost-gluon 
vertex to the following forms; 
\begin{eqnarray}
V_{\mu\nu\rho}^{abc}(p,q,r;\Lambda)&=&0,
\label{18}\\
T_\mu^{abc}(p|q,r;\Lambda)&=&ig(\Lambda)f^{abc}q_\mu,
\label{19}
\end{eqnarray}
where the three gluon vertex is just neglected.
It is because the gluon loop corrections give sub-leading contributions
concerning to the infrared momentum region where the power behavior is 
seen in the SDE 
\cite{smekal_1, smekal_2, fischer_1, fischer_2, zwanziger_1}.
Of course the three gluon vertex is not negligible at ultraviolet 
and intermediate momentum region, and therefore resultant RG flows
cannot be regarded as those for the Yang-Mills theories.
However we examine the ERG by imposing this condition just for simplicity
as the first step of the analyses.
While the three ghost-gluon vertex is given by the bare form (\ref{19}). 
Owing to the non-renormalization theorem for the three ghost-gluon 
vertex in Landau gauge\cite{taylor}, the coupling constant is not
modified $g(\Lambda)=g(\Lambda_0)$, 
In fact, if we truncate the bare form (\ref{19}), 
the coupling non-renormalization $\partial_\Lambda g(\Lambda)=0$ 
is also shown by using the flow equation
similarly to the perturbative one-loop argument. 

\begin{figure}
\centerline{\includegraphics[width=12cm]{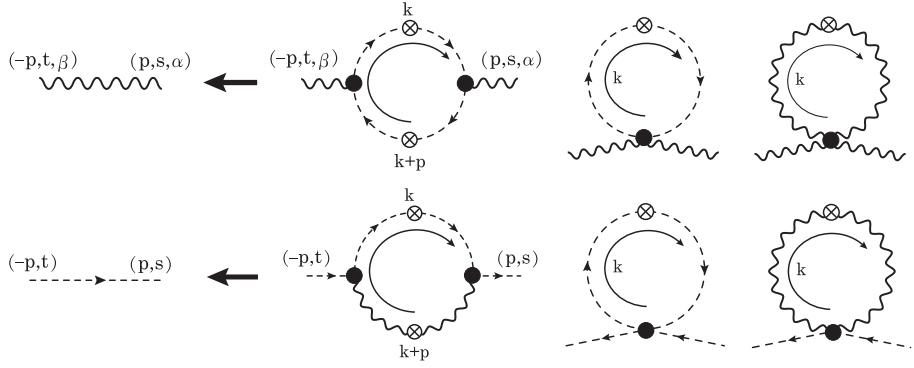}}
\caption{The diagrammatic representation of the beta 
functions for the gluon and the ghost two point functions; 
$f_Z(p^2;\Lambda)$ (upper diagram) and $f_G(p;\Lambda)$ (lower diagram). 
The crossed circle implies insertion of the cutoff function 
$\partial_\Lambda R_\Lambda $.}
\label{fig1}
\end{figure}

Now let us derive the flow equations for the two point functions.
From now on we will consider SU(3) Yang-Mills theory, although the extension 
to general SU(N) gauge group can be done without difficulty.  
Contracting the flow equation for the gluon two point function 
with BP tensor $B_{\mu\nu}(p)$,
we can easily obtain the following flow equation,
\begin{eqnarray}
&&\frac{1}{2}\delta^{ab}p^2\partial_\Lambda f_Z(p^2;\Lambda)\nonumber\\
&=&-\frac{1}{2}\delta^{ab}g^2(\Lambda_0)
\int_k\,k^2\partial_\Lambda R_\Lambda(k^2)\,
P_G^2(k^2;\Lambda) P_G\left((k+p)^2;\Lambda\right)\,
\left[k_\alpha (k+p)_\beta B_{\alpha\beta}(p)\right]\nonumber\\
&&-\frac{1}{2}\delta^{ab}g^2(\Lambda_0)
\int_k\,(k+p)^2\partial_\Lambda R_\Lambda\left((k+p)^2\right)\,
P_G^2\left((k+p)^2;\Lambda\right) P_G\left(k^2;\Lambda\right)\,
\left[k_\alpha (k+p)_\beta B_{\alpha\beta}(p)\right]\nonumber\\
&&-\frac{1}{3}\int_k\,k^2\partial_\Lambda R_\Lambda(k^2)\,
P_G^2\left(k^2;\Lambda\right)\,
\left[B_{\alpha\beta}(p)T_{\alpha\beta}^{abcd}
\left(p,-p|k,-k;\Lambda\right)\delta^{cd}\right]\nonumber\\
&&-\frac{1}{6}\int_k\,k^2\partial_\Lambda R_\Lambda(k^2)\,
P_Z^2\left(k^2;\Lambda\right)\,
\left[B_{\alpha\beta}(p)V_{\alpha\beta\mu\nu}^{abcd}(p,-p,k,-k;\Lambda)
T_{\mu\nu}(k)\delta^{cd}\right]\times 12,
\label{20}
\end{eqnarray}
for the ghost two point function.
The flow equation for the gluon two point function is 
also found to be
\begin{eqnarray}
&&-\delta^{ab}p^2\partial_\Lambda 
f_G(p^2;\Lambda)\nonumber\\
&=&-3\delta^{ab}g^2(\Lambda_0)\int_k\,k^2\partial_\Lambda 
R_\Lambda(k^2)\,P_G^2\left(k^2;\Lambda\right) 
P_Z\left((k+p)^2;\Lambda\right)\,
\left[k_\mu k_\nu T_{\mu\nu}(k+p)\right]\nonumber\\
&&-3\delta^{ab}g^2(\Lambda_0)\int_k\,(k+p)^2
\partial_\Lambda R_\Lambda\left((k+p)^2\right)\,
P_Z^2\left((k+p)^2;\Lambda\right) 
P_G\left(k^2;\Lambda\right)\,
\left[k_\mu k_\nu T_{\mu\nu}(k+p)\right]\nonumber\\
&&-\int_k\,k^2\partial_\Lambda R_\Lambda(k^2)\,
P_G^2\left(k^2;\Lambda\right)\,
\left[T^{abcd}\left(p,-p,k,-k;\Lambda\right)
\delta^{cd}\right]\times 4
\nonumber\\
&&-\frac{1}{2}\int_k\,k^2
\partial_\Lambda R_\Lambda(k^2)\,
P_Z^2\left(k^2;\Lambda\right)\,
\left[T_{\mu\nu}(k) T_{\mu\nu}^{cdab}\left(k,-k|p,-p;\Lambda\right)
\delta^{cd}\right]\times 2,
\label{21}
\end{eqnarray}
where we introduced short hand notations for the cutoff propagators; 
\begin{eqnarray}
P_Z\left(p^2;\Lambda\right)&=&
\frac{1}{k^2 [f_Z(k^2;\Lambda)+R_\Lambda(k^2)]},
\label{22}\\
P_G\left(p^2;\Lambda\right)&=&
\frac{1}{k^2 [f_G(k^2;\Lambda)+R_\Lambda(k^2)]}.
\label{23}
\end{eqnarray}
The integrals in the beta functions may be interpreted as specific
one loop corrections and their diagrammatic representations are 
presented in Fig.\ref{fig1}. 

Even though the three point vertices are simple in the Landau gauge, 
it is still difficult to solve the flow equations. 
Because the two point functions 
and the four point functions appearing in the beta functions
are {\it fully momentum dependent}. 
In oder to integrate the flow equations toward infrared numerically, 
it is quite preferable to perform the shell momentum integrals 
in the RHS of the flow equations analytically. 
Then the numerical analyses are much simplified.
For this purpose, we adopt one dimensional (y-max) approximation and 
take sharp cutoff limit as follows. 

The one dimensional approximation is often used also in SDE analyses, 
which replaces the each form factor as, 
\begin{equation}
f_i\left((k+p)^2;\Lambda\right)\rightarrow 
f_i\left({\rm max}\left\{k^2,p^2\right\};\Lambda\right),\quad
(i=G,Z),
\label{24}
\end{equation} 
in the above flow equations (\ref{20}) and (\ref{21}). 
Next we consider to perform the angular integrals of internal momentum.
It is noted that the cutoff propagators $P_i\left((k+p)^2;\Lambda\right)$
are reduced to the factorized form as
\begin{equation}
P_i\left((k+p)^2;\Lambda\right)
\rightarrow\frac{1}{(k+p)^2f_i\left({\rm max}
\left\{k^2,p^2\right\};\Lambda\right)}
\theta^{\epsilon}\left((k+p)^2-\Lambda^2\right),
\label{25}
\end{equation}
in the sharp cutoff limit given by (\ref{13}).  
The superscript $\epsilon$ indicates the sharpness of the cutoff 
function ($\epsilon\sim\alpha^{-1}$ for (\ref{13})). 
However, naive sharp cutoff limit induces non-analyticity at the origin 
of momentum space, reflecting the non-local behavior in position space.
One way out this problem is to use momentum scale 
$p\sim\sqrt{p_\mu p_\mu}$ instead of the momentum components 
$p_\mu$\cite{morris_1}. 
Here, however, we avoid this problem by simply replacing the all 
momenta associated with the cutoff propagators $k+p_1+\cdots +p_n$ 
($k$ is inner-loop momentum, and $p_i$ is external momentum) with 
the inner-loop momentum $k$. 
Resultantly the above manipulations correspond to replacement of the 
cutoff propagators as
\begin{equation}
P_i\left((k+p)^2;\Lambda\right)\rightarrow
\frac{1}{(k+p)^2f_i\left({\rm max}
\left\{k^2,p^2\right\};\Lambda\right)}
\theta^\epsilon\left(k^2-\Lambda^2\right).
\label{26}
\end{equation}
The $\Lambda$ derivative of the cutoff propagators are also
replaced by 
\begin{eqnarray}
&&(k+p)^2\partial_\Lambda R_\Lambda
\left((k+p)^2\right)\,P_i\left((k+p)^2;\Lambda\right)\nonumber\\
&=&-\partial_\Lambda\left\{\frac{1}{(k+p)^2
\left[f_i\left((k+p)^2;\Lambda\right)+R_\Lambda\left((k+p)^2\right)\right]}
\right\}_{\Lambda'=\Lambda}\nonumber\\
&\rightarrow&-\frac{1}{(k+p)^2f_i
\left({\rm max}\left\{k^2,p^2\right\};\Lambda\right)}
\partial_\Lambda\theta^\epsilon\left(k^2-\Lambda^2\right).
\label{27}
\end{eqnarray}
After this replacement, the shell integrals may be carried out 
analytically by using the formula \cite{morris_1}; 
\begin{equation}
\delta^\epsilon\left(k^2-\Lambda^2\right)\,h
\left(\theta^\epsilon(k^2-\Lambda^2)\right)
\rightarrow\delta\left(k^2-\Lambda^2\right)
\int^1_0 dt h\left(t\right)\quad 
{\rm as}\quad \epsilon\rightarrow0,
\label{28}
\end{equation}
where $h\left(\theta^\epsilon(k^2-\Lambda^2)\right)$ is an 
arbitrary function of the step function $\theta^{\epsilon}$. 
Indeed it is far from obvious whether the one dimensional 
approximation and the sharp cutoff limit introduced in our approximation
scheme is effective well or not. However it would be inevitable 
to use of these approximations in order to solve the flow equations with 
retaining the full momentum dependence of the two and four point 
vertex functions. Needless to say this approximation may be too
brute, although the analysis of the flow equations becomes easy.
Another approach base on derivative expansion of the effective 
average action may be also possible, however, we would like to 
leave it for future work\cite{jkato}. 

\begin{figure}
\centerline{\includegraphics[width=13cm]{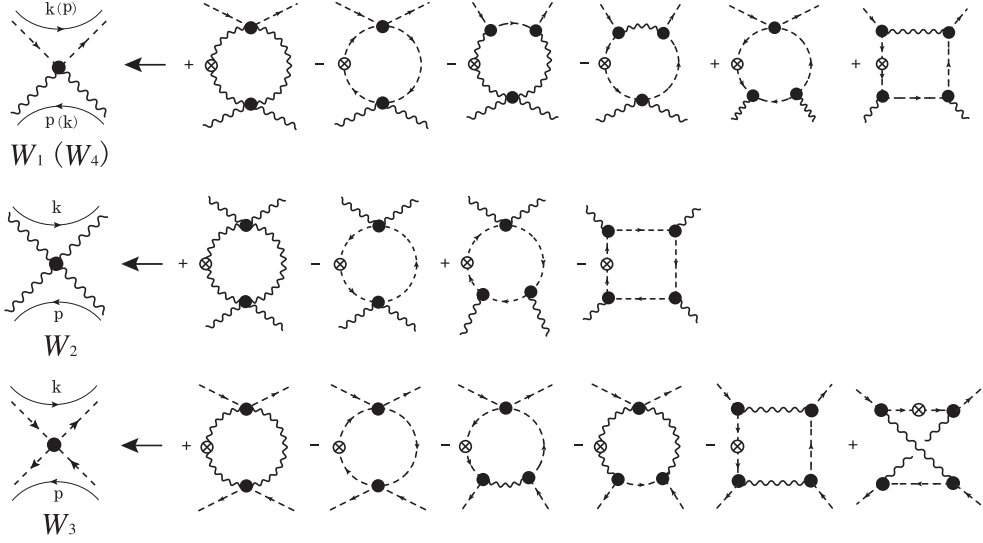}}
\caption{Diagrammatic representations of the beta functions for the 
four point vertex functions: 
the 4-ghost-gluon vertices $W_1(p^2,k^2;\Lambda)$ and 
$W_4(p^2,k^2;\Lambda)$ (upper line), 
the 4-gluon vertex $W_2(p^2,k^2;\Lambda)$ (center line) and 
the 4-ghost vertex $W_3(p^2,k^2;\Lambda)$ (lower line). 
Here $p$ denotes the external and $k$ denotes the internal momenta 
when they are used in the flow equations for the two point 
functions shown in Fig.\ref{fig1}.} 
\label{fig2}
\end{figure}

Lastly let us consider the four point vertices. 
By substituting the truncated action (\ref{14}) into the flow 
equation (\ref{6}), we may obtain the beta functions for the four 
point vertices: 
$T_{\mu\nu}^{abcd}(p,q|r,s;\Lambda), 
V_{\mu\nu\rho\sigma}^{abcd}(p,q,r,s;\Lambda), 
T^{abcd}(p,q,r,s;\Lambda)$. 
However it is still difficult to solve these flow equations with retaining 
all the informations of the three independent momenta, color and Lorentz 
indices.
Therefore we shall further restrict their functional forms below. 
As stated at the beginning of this section, our truncation scheme 
deals with only the diagrams which contribute to the two point 
functions. 
Looking at the flow equations (\ref{20}) and (\ref{21}), we observe 
that the four types of four point vertices appear in the flow equations
with two independent momenta (the external momentum $p$ and the internal
momentum $k$) associated with the following momentum channels; 
\begin{eqnarray}
T_{\alpha\beta}^{abcd}(p,-p|k,-k;\Lambda),&&
\quad V_{\mu\nu\alpha\beta}^{abcd}(p,-p,k,-k;\Lambda)\label{29},\\
T^{abcd}(p,-p,k,-k;\Lambda),&&
\quad T_{\mu\nu}^{cdab}(k,-k|p,-p;\Lambda).
\label{30}
\end{eqnarray}
Here, we have distinguished the internal momentum $k$ from the external
momentum $p$, because these momenta are treated asymmetrically in the 
followings.
Then we shall approximate the beta functions of the four 
point vertices by restricting to the above types of vertices;
we neglect differences in the other momentum channels.  
Also we restrict the color and Lorentz index structures with 
the following forms; 
\begin{eqnarray}
T_{\alpha\beta}^{abcd}(p,-p|k,-k;\Lambda)&=&
\left[\delta^{ab}\delta^{cd}/8\right]\,
\left[T_{\alpha\beta}(p)/3\right]\,
W_1\left(p^2,k^2;\Lambda\right),
\label{31}\\
V_{\mu\nu\alpha\beta}^{abcd}(p,-p,k,-k;\Lambda)&=&
\left[\delta^{ab}\delta^{cd}/8\right]\,
\left[T_{\alpha\beta}(p)/3\times T_{\mu\nu}(k)/3\right]\,
W_2\left(p^2,k^2;\Lambda\right),
\label{32}\\
T^{abcd}(p,-p,k,-k;\Lambda)&=&
\left[\delta^{ab}\delta^{cd}/8\right]\,
W_3\left(p^2,k^2;\Lambda\right),
\label{33}\\
T_{\mu\nu}^{cdab}(k,-k|p,-p;\Lambda)&=&
\left[\delta^{ab}\delta^{cd}/8\right]\,
\left[T_{\mu\nu}(k)/3\right]\,
W_4\left(p^2,k^2;\Lambda\right),
\label{34}
\end{eqnarray}
where $T_{\mu\nu}(p)$ denotes the transverse projector and
the normalization factors are chosen for the later conveniences.
Here we introduced the four types of four point vertices 
$W_1$, $W_2$, $W_3$, $W_4$ as functions dependent on two momentum 
variables $p^2$ and $k^2$ as well as the cutoff scale $\Lambda$.
Thus we have reduced the four point vertices to these scalar 
functions by restricting structures of the four point vertices, and also
by using the one dimensional approximation in the sharp cutoff limit.
Then the flow equations for the four point vertices are reduced to 
a set of coupled partial differential equations with respect to the 
momentum dependent functions: $f_G, f_Z, W_1, W_2, W_3, W_4$.
One may wonder if the functions $W_1$ and $W_4$ are not
independent mutually, as is seen from their definition given
by (\ref{31}) and (\ref{34}). 
However we will treat them as independent functions
because of the following reason in approximation.
In order to write down the beta functions for $W_i$, it is
necessary to evaluate the box diagrams which consist of the 
three ghost-gluon vertices (see Fig.\ref{fig2}). 
Then one encounters the two angle integrations
\footnote{There are only a single angle integration in other diagrams, 
because $W_i(p^2,k^2;\Lambda)$ are functions of
$p^2$ and $k^2$.}, 
which are induced by the propagtors associated with the momenta 
$(k'+p)^2$ and $(k'+k)^2$, where $k'$ is the internal loop momentum 
in the box diagrams.
This angular integrals in the box diagrams cannot be performed 
analytically.
Here, therefore, we will approximate the integrand by expanding the 
inner products of momenta $k_\mu$, that is, $(k'k)$ and $(pk)$
to the second order, and then average over the direction of $k_\mu$.
Thus $W_1$ and $W_4$ receive different radiative corrections 
in this evaluation of beta functions, so 
we treat them as independent functions.

Resultantly, our ERG equations are reduced to a set of coupled partial 
differential equations for the two propagator functions 
$f_Z(p^2;\Lambda)$, $f_G(p^2;\Lambda)$ and four types of the scalar 
functions $W_i(p^2,k^2;\Lambda),~(i=1,\cdots,4)$.
Diagrammatic representations of the beta functions for the two and
four point functions are depicted in Fig.\ref{fig1} and Fig.\ref{fig2}
respectively.
Concrete expressions of the flow equations are summarized in Appendix.A. 
We will solve this set of ERG equations numerically in the next section. 

Before closing this section it is also helpful to define more primitive 
approximation scheme for later discussions. 
We define the scheme in which the four point vertices are neglected by putting  
$W_i(p^2,k^2;\Lambda)=0$ under the RG evolution. 
This scheme is equivalent to take into account only the first diagrams 
of each flow equations for $f_Z(p^2;\Lambda)$ and $ f_G(p^2;\Lambda)$ 
given in Fig.\ref{fig1}. 
At first sight this scheme is more similar to the SDE approximation 
scheme, which has the infrared power solution. 
However, as we will show in the next section, this scheme will fail to 
reach deep infrared region. 
The reason will be considered in the last section. 

\section{Results of numerical analysis\label{result}}

In this section we will present the results obtained by numerically 
integrating the flow equations, specially with paying much attention on
infrared behavior of the gluon and the ghost propagator functions,
$f_Z(p^2;\Lambda)$ and $f_G(p^2;\Lambda)$. 
First of all, we need to specify the initial condition for the 
effective action in order to proceed the numerical integration. 
If one choose a very large scale as the initial cutoff 
scale $\Lambda_0$, then the boundary condition for the flow equations
may well be given by the bare action.
This is because the arbitrary irrelevant terms do not affect the
infrared behavior according to the discussion given in section
\ref{exactrenormalizationgroup}. 
Obviously the large initial cutoff scale $\Lambda_0$ is 
characterized by the small coupling constant in asymptotically free 
theories. 
We thus define the initial conditions at the initial cutoff scale 
$\Lambda_0$ simply by 
\begin{eqnarray}
&&g(\Lambda_0)=2.0,
\label{35}\\
&&f_Z(p^2;\Lambda_0) = f_G(p^2;\Lambda_0) = 1,
\label{36}\\
&&W_i(p^2,k^2;\Lambda_0) = 0 \quad(i=1,\cdots,4).
\label{37}
\end{eqnarray}
Here it should be noted also that the bare four gluon vertex does not 
appear in the initial conditions (\ref{37}). 
The presence of such a momentum independent four point vertex 
is irrelevant in the flow equations, because the corrections
are contracted by the B-P tensor $B_{\mu\nu}(p)$.

It would be better to use the initial conditions defined with
including quantum fluctuation from larger momentum region, $p^2>\Lambda_0^2$. 
This is easily carried out by solving the perturbative flow equations 
under suitable renormalization conditions\cite{ellwanger_1, bergerhoff}. 
However such corrections are relevant for flows only at the 
ultraviolet region. 
Our main concern is to see whether the infrared power solution is 
realized as an infrared attractive solution in the ERG framework.
Therefore we employ the above simple initial conditions 
(\ref{35})$\sim$(\ref{37}). 

\begin{figure}[t]
\centerline{\includegraphics[width=12cm]{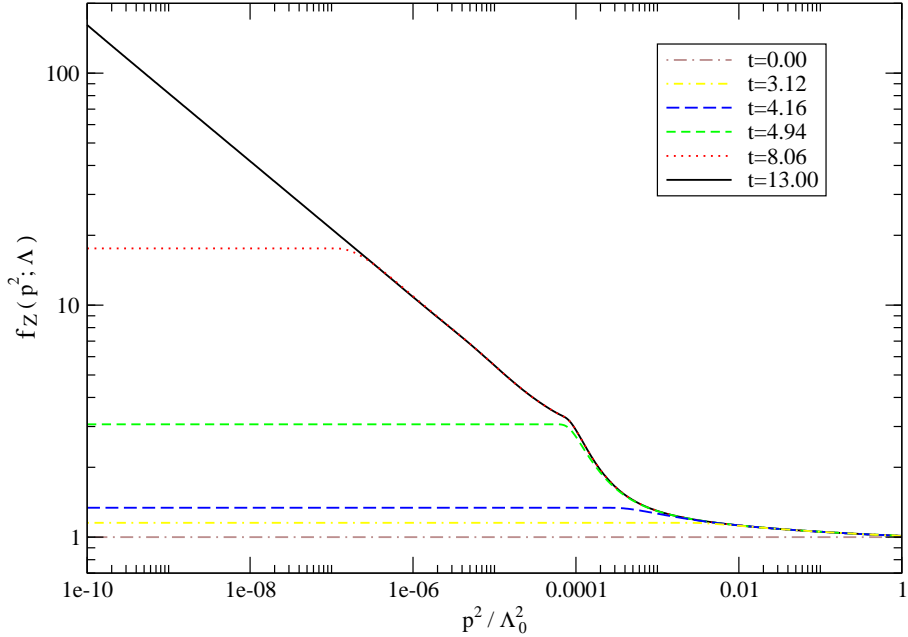}}
\caption{The flow ($\Lambda$ dependence) of the gluon propagator 
function $f_Z(p^2;\Lambda)$ for various cutoff scales of
$t={\rm log}(\Lambda_0/\Lambda)=
0.00, 3.12, 4.16, 4.94, 8.06, 13.00$. } 
\label{fig3}
\end{figure}

\begin{figure}[thb]
\centerline{\includegraphics[width=12cm]{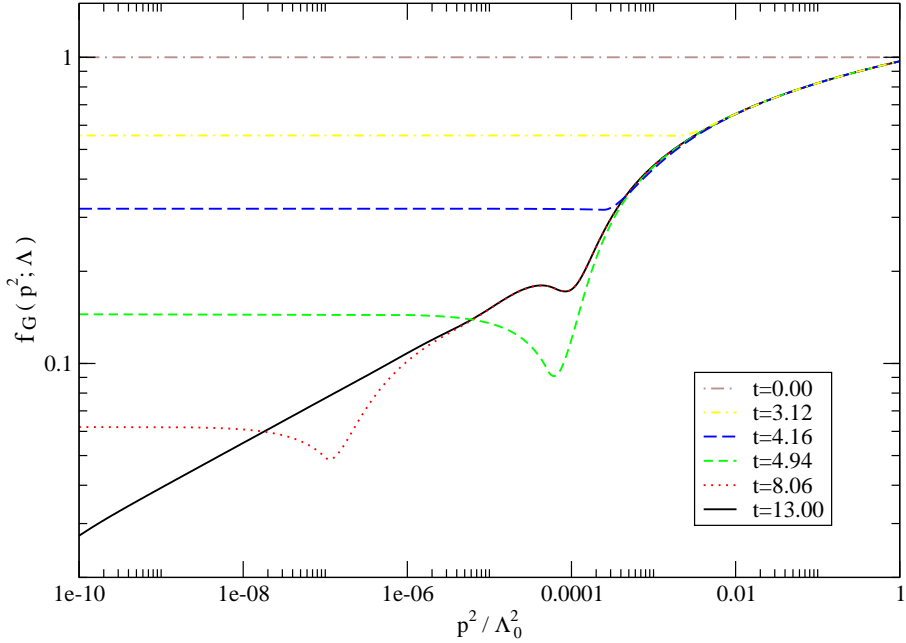}}
\caption{The flows ($\Lambda$ dependence) of the ghost propagator 
function $f_G(p^2;\Lambda)$ at various cutoff scales 
$t=0.00, 3.12, 4.16, 4.94, 8.06, 13.00$.} 
\label{fig4}
\end{figure}

Now let us present the results of numerical analysis of the flow equations.
In Figs.\ref{fig3} and \ref{fig4}, we show flows of the gluon function 
$f_Z(p^2;\Lambda)$ and the ghost function 
$f_G(p^2;\Lambda)$ respectively with varying the 
infrared cutoff scale $\Lambda =\exp(-t)\Lambda_0$. 
In the figures their snapshots at 
$t=0.00, 3.12, 4.16, 4.94, 8.06, 13.00$ are presented in one panel. 
%The $p^2$-horizontal axis is given in units of $\Lambda_0^2$ and graphs are plo%tted in double log scale. 
{}From these flows one may observe that the propagator functions 
$f_Z(p^2;\Lambda)$ and $f_G(p^2;\Lambda)$ freeze immediately 
once the cutoff scale is lowered than the momentum scale, namely, at
the region of $\Lambda^2< p^2 <\Lambda^2_0$. 
The graphs are strongly bended around $p^2 \sim \Lambda^2$
and become almost flat at lower momentum region. 
It may be supposed that the bends are due to the crude
approximation. Because the integrand of these beta functions are 
infrared cutoff, and therefore the corrections around the 
$p^2 \sim \Lambda^2$
may be overestimated due to the one dimensional approximation
in the sharp cutoff limit employed in the previous section. 
Anyway, as lowering the cutoff scale $\Lambda$, it is found 
that the frozen 
solutions drastically change around $p^2/\Lambda_0^2\sim 0.0001$ and 
eventually show the power scaling. 
This means also that power behavior is certainly realized 
in the infrared propagators defined by (\ref{17}).

\begin{figure}[thb]
\centerline{\includegraphics[width=12cm]{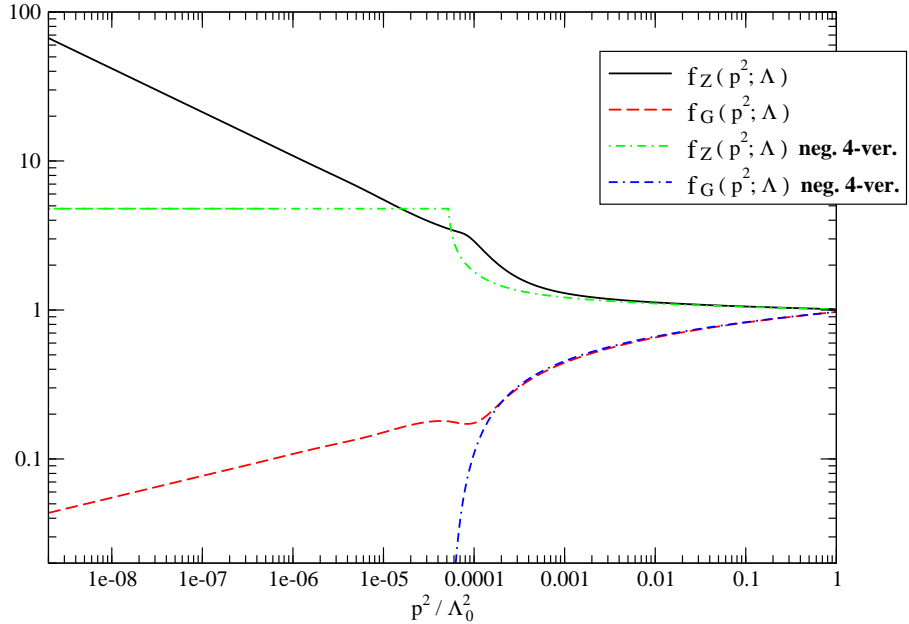}}
\caption{The propagator functions $f_Z(p^2;\Lambda)$ (full line) and 
$f_G(p^2;\Lambda)$ (long-dashed line) obtained at $t=13.00$,
which show power behavior in the infrared region. 
For comparison these functions evaluated with neglecting the four 
point vertices are also shown for 
$f_Z(p^2;\Lambda)$ (dashed-dotted line) and $f_G(p^2;\Lambda)$ 
(dashed-dashed-dotted line) just before reaching singularity 
(t=4.93).}
\label{fig5}
\end{figure}

In Fig.\ref{fig5} the result for these propagator functions
obtained at $t=13.00$ are shown. There we also show these functions
evaluated with neglecting the four point vertices $W_i(p^2,k^2;\Lambda)$ 
in the flow equations for comparison.
It is seen clearly that these functions enjoy power behavior
in the infrared without singular poles. 
On the other hand if the four point vertices are neglected, 
we observe that the ghost propagator function diverges at a finite 
scale, and our numerical calculation is broken down. In the
present case, the singularity appears at $t\approx5.00$. 
The snapshots in Fig.\ref{fig5} is given adjacent to the singularity. 
It is seen that the two schemes present qualitatively the same behavior 
at the ultraviolet region, however the evolutions in the infrared become
drastically different. Thus we may conclude that  contributions through
the four point vertices are significant in the infrared region to 
achieve power behavior with avoiding singularity. 

\begin{figure}[htb]
\centerline{\includegraphics[width=12cm]{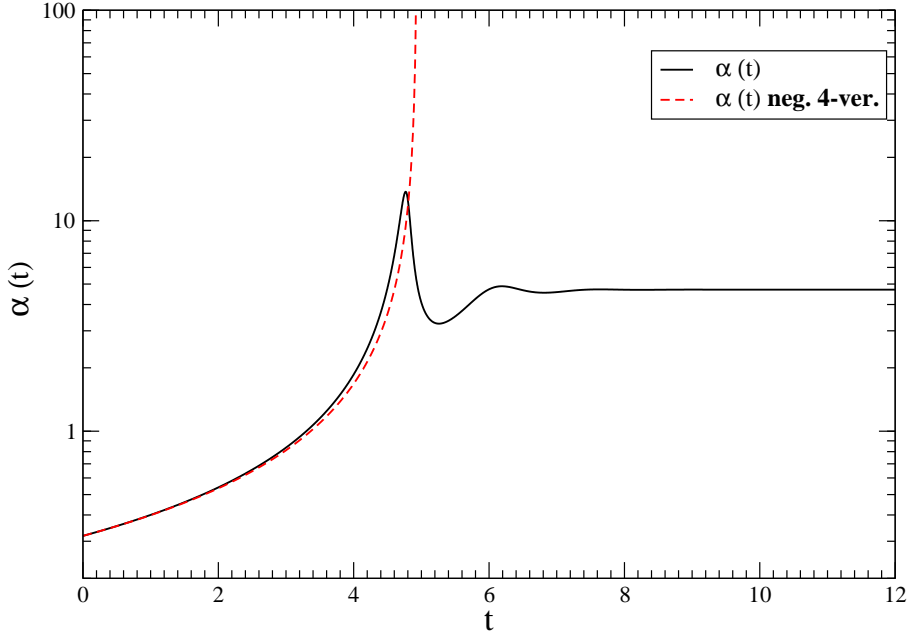}}
\caption{The running coupling $\alpha(t)$ (solid line). 
The long-dashed line stands for the running coupling obtained neglecting
four point vertices. } 
\label{fig6}
\end{figure}

Interestingly these infrared power solutions are related with the 
running gauge coupling constant, which may be defined naively
in the RG framework by 
\begin{equation}
\alpha(t) = 
\frac{g^2(\Lambda_0)}{4\pi f_Z(p^2=0;\Lambda(t))f_G^2(p^2=0;\Lambda(t))}.
\label{38}
\end{equation}
The running coupling obtained by the present analysis is shown 
in Fig.\ref{fig6} compared with that in the scheme neglecting the
four vertices. 
We see that the running coupling converges to an infrared fixed point 
$\alpha^*\approx 4.70$.
In contrast to this the running coupling without four vertex contributions
is found to hit the Landau pole.
In the SDE approach, the running gauge coupling is sometimes defined
as a function of momentum, which is regarded as the renormalization scale \cite{smekal_1, smekal_2}.
In our ERG framework, the coupling constant would correspond to
\begin{equation}
\alpha(p^2) = 
\frac{g^2(\Lambda_0)}{4\pi f_Z(p^2;\Lambda \rightarrow 0)
f_G^2(p^2;\Lambda \rightarrow 0)}.
\end{equation}
It is seen also that this coupling constant shows the qualitatively
same behavior with $\alpha(t)$ and that an infrared fixed 
point exists.

{}From the definition of the running coupling (\ref{38}), 
it is clear that appearance of the infrared fixed point can be 
realized by exact cancellation of the corrections to 
the gluon and ghost propagators.
The infrared power behavior implies that the exponents must
satisfy the specific ratio as shown in (\ref{1}). 
The exponents, that is the anomalous dimensions, may be 
defined also as $\Lambda$-dependent quantity,  
\begin{equation}
\eta_Z(t) = -\frac{1}{2}\frac{\partial}{\partial t} 
{\rm log}f_Z(p^2=0;\Lambda(t)), \quad \quad
\eta_G(t) = -\frac{1}{2}\frac{\partial}{\partial t} 
{\rm log}f_G(p^2=0;\Lambda(t)).
\label{39}
\end{equation}
In the infrared limit $(t=13)$ these exponents are found to converge to
stable constants given by  
\begin{equation}
\kappa \equiv \eta_G^* = 
-\frac{1}{2}\eta_Z^* \approx 0.146,
\label{40}
\end{equation}
where $\kappa$ denotes the exponent in Eq.~(\ref{1})  used frequently
in the SDE studies .

\section{Conclusion and discussion\label{conclusionanddisscussion}}
We have performed numerical analysis of the approximated ERG equations 
for Landau gauge SU(3) Yang-Mills theory. There flows of the momentum 
dependent four point vertices as well as the momentum dependent gluon 
and ghost propagators were examined. 
In order to setup the approximated flow equations with retaining momentum 
dependence of vertices in a minimal way, we restricted the diagrams to
those which contribute to the two point functions.
Further we employed one-dimensional approximation and sharp cutoff limit
to make numerical analysis easier. 
The resultant flow equations are given in the form of coupled partial 
differential equations which have been solved numerically.
We solved the flow equations with an ordinary bare action as the initial
conditions for the effective action.
However infrared behavior of the flows are not affected by the
arbitrary irrelevant terms in the effective action at ultraviolet.

Our numerical solution exhibits infrared power behavior for the propagators
$D_Z(p^2)\sim(p^2)^{-1+2\kappa}\,,D_G(p^2)\sim(p^2)^{-1-\kappa}$
with the exponent $\kappa\approx0.146$, which is similar to the result
obtained by SDE studies. 
Also the running gauge coupling is found to approach an infrared fixed 
point $\alpha^*\approx4.70$.
Compared with the SDE analyses, the most different point is that 
inclusion of the four point vertices generated during flow 
is crucial to obtain the power solutions with avoiding infrared divergence. 
In the SDE approach, only the dressed four gluon vertex appears through 
the two loop diagram in the equation for gluon propagator and existence 
of the power solution has been confirmed irrespectively to the 
vertex contributions
\cite{smekal_1, smekal_2, fischer_1, fischer_2,atkinson_1,kondo_2, bloch_1}. 

This difference may be considered as follows. 
The SDEs are a set of integral equations among the Green functions, 
and there {\it all momentum modes} are coupled to each other. 
If we solve the SDEs for the propagators iteratively, then the
propagotor may be represented by an infinite sum of diagrams whose
internal momenta are not restricted.
On the other hand the ERG equation is reduced to a set of the 
beta functions for the Green functions, which are given
by one-loop exact forms diagrammatically. 
Therefore if we could solve the flow equations iteratively, then the
Green functions are given by an infinite sum of diagrams involving
all loops, which must be identical to the solution of the SDE.
However this is not the case after truncation.
Note that internal momenta of the one-loop integrals of the beta
functions are restricted around the cutoff scale $\Lambda$, and 
contributions from higher momentum loops are already taken in the 
effective vertices in the ERG framework.
Also the beta functions for the effective vertices involve the
higher point vertices as well as the lower ones.
Therefore, in our present concern, some part of the corrections to 
the two point functions comes through four point effective vertices.
If we truncate the effective action and discard the four and higher point
functions, the corrections generated through higher point vertices
are lost in the resultant propagators.
Speaking more explicitly, the diagrams in which the inner loops carry
lower momenta than the outer loops are suppressed in the diagrammatic
terminology.
Inclusion of the four point functions improve this situation
drastically indeed.
In the case of Landau gauge Yang-Mills theory, the ghost 
propagator is considered to be enhanced at infrared. 
Therefore, particularly couplings of the soft ghost modes with 
other high energy modes 
are expected to give non-negligible contributions.

Finally we shall discuss residual problems and the advantageous 
features of the ERG approach. 
Although we found the power solution in the ERG framework, the obtained 
exponent $\kappa\approx0.146$ is fairly smaller than the currently 
most reliable value $\kappa\approx0.5$ expected
from the SDE analysis and also from axiomatic considerations. 
Furthermore transient from the logarithmic to the power 
behavior of the propagators is not smooth.
%the obtained running coupling is not monotonic function of
%the momentum scale, therefore this lead to un-physical 
%multi-valued beta function. 
Probably these are due to the fact that our approximation scheme 
is  still poor in the following points.
Firstly the corrections for the two point functions with 
$p^2 \sim \Lambda^2$ are fairly influenced by use of the one-dimensional
approximation and the sharp cutoff limit. 
Then the four point vertices which have other channels and also
higher order vertices may contribute to flows of the two point 
functions.
These issues are currently investigated with employing another type of
approximation scheme\cite{jkato}. 
Further, inclusion of three gluon interactions, which have been omitted
in this paper just for the simplicity, and also dynamical quarks
are remained as obvious problems.

The advantageous  features of ERG may be considered as follows.
We could obtain the infrared power behavior for the gluon and the ghost
propagators without any ansatz for their infrared behaviors, and by
starting from rather arbitrary ultraviolet bare actions.
On the contrary all the SDE analyses so far were based on the infrared 
asymptotic analysis, which needs an appropriate ansatz. 
In the case of Landau gauge Yang-Mills theories, the power law ansatz
was found to suit nicely. However in the other situations such as 
in the deconfinement phase the appropriate ansatz is unknown a priori. 
It would be another advantage of the ERG to obtain some informations
of the  infrared four point functions at the same time.
In these respects the ERG formalism seems to be suitable to explore 
the infrared dynamics in more general cases. Especially exploration of
dynamics near phase transitions would be quite interesting and 
challenging problem to be considered, and the ERG approach is
expected to be useful also in such studies.

Note added : After completion of this paper, Ref.\cite{pawlowski}
appeared which addresses related issues.

\begin{acknowledgments}
I would like to thank H. Terao for helpful discussions and for reading 
the manuscript. 
I would also like to thank K.-I. Aoki, T. Izubuchi 
and K. Takagi for useful discussions. 
\end{acknowledgments}

\appendix
\section{approximated flow equations\label{flowequations}}
The flow equations derived by using the approximation scheme 
explained in section \ref{approximationscheme} are summarized here. 
The all momentum variables and dimensionful functions are 
normalized by the cutoff scale $\Lambda$, and we use
$x=p^2/\Lambda^2, y=k^2/\Lambda^2$. 
By contracting the external gluon momentum with B-P tensor 
$B_{\mu\nu}(p)$, all vertex functions $W_i (x,y;\Lambda)$ are 
factorized with the momentum variables. 
So we define the factorized functions by
$W_i(x,y;\Lambda) = x y \times\bar{W}_i(x,y;\Lambda)$
and use them below. 
The flow equations, which are somewhat lengthy though, 
are found to be in the followings.

\begin{eqnarray}
\partial_t f_Z(x;\Lambda)&=&-2x\partial_x f_Z(x;\Lambda)
\nonumber\\
&&+\frac{g^2(\Lambda_0)}{16\pi^2} f_G^{-1}(1;\Lambda)
f_G^{-1}({\rm max}(x,1);\Lambda)\times
\left\{\begin{array}{ll}x^{-2}(3-2x^{-1})&(x>1)\\1&(x\le1)\end{array}
\right.
\nonumber\\
&&+\frac{1}{12\pi^2}f_G^{-1}(1;\Lambda)\,\bar{W}_1(x,1;\Lambda)
+\frac{1}{2\pi^2}f_Z^{-1}(1;\Lambda)\,\bar{W}_2(x,1;\Lambda),
\label{a1}\\
\partial_t f_G(x;\Lambda)&=&-2x\partial_xf_G(x;\Lambda)
\nonumber\\
&&-\frac{g^2(\Lambda_0)}{32\pi^2}f_G^{-1}(1;\Lambda)
f_Z^{-1}({\rm max}(x,1);\Lambda)\times
\left\{\begin{array}{ll}x^{-2}&(x>1)\\1&(x\le1)\end{array}\right.
\nonumber\\
&&-\frac{1}{2\pi^2}f_G^{-1}(1;\Lambda)\,\bar{W}_3(x,1;\Lambda)
-\frac{1}{8\pi^2}f_Z^{-1}(1;\Lambda)\,\bar{W}_4(x,1;\Lambda),
\label{a2}
\end{eqnarray}
\begin{eqnarray}
&&\partial_t \bar{W}_1(x,y;\Lambda)
\nonumber\\
&=&-4\bar{W}_1(x,y;\Lambda)-2x\partial_x\bar{W}_1(x,y;\Lambda)
-2y\partial_y\bar{W}_1(x,y;\Lambda)
\nonumber\\
&&-\frac{1}{2\pi^2}f_Z^{-2}(1;\Lambda)\,
\bar{W}_2(x,1;\Lambda)\bar{W}_4(y,1;\Lambda)
+\frac{1}{2\pi^2}f_G^{-2}(1;\Lambda)\,
\bar{W}_1(x,1;\Lambda)\bar{W}_3(y,1;\Lambda)
\nonumber\\
&&-\frac{3g^2(\Lambda_0)}{8\pi^2}f_Z^{-2}(1;\Lambda)
f_G^{-1}({\rm max}(y,1);\Lambda)\,
\bar{W}_2(x,1;\Lambda)\times
\left\{\begin{array}{ll}(3y^{-1}-y^{-2})&(y>1)\\(3-y)&(y\le1)\end{array}
\right.
\nonumber\\
&&+\frac{9g^2(\Lambda_0)}{32\pi^2}f_G^{-2}(1;\Lambda)
f_Z^{-1}({\rm max}(y,1);\Lambda)\,\bar{W}_1(x,1;\Lambda)\times
\left\{\begin{array}{ll}y^{-2}&(y>1)\\1&(y\le1)\end{array}
\right.
\nonumber\\
&&+\frac{3g^2(\Lambda_0)}{4\pi^2}f_G^{-2}(1;\Lambda)
f_G^{-1}({\rm max}(x,1);\Lambda)\,\bar{W}_3(y,1;\Lambda)\times
\left\{\begin{array}{ll}(3x^{-2}-2x^{-3})&(x>1)\\1&(x\le1)\end{array}
\right.
\nonumber\\
&&+\frac{3g^4(\Lambda_0)}{8\pi^2}f_G^{-2}(1;\Lambda)
f_G^{-1}({\rm max}(x,1);\Lambda)f_Z^{-1}({\rm max}(y,1);\Lambda)\,
\nonumber\\
&&\quad\times\left(\frac{(1+y(6+y))}{(1+y)^4}\right)\times
\left\{\begin{array}{ll}(3x^{-2}-2x^{-3})&(x>1)\\1&(x\le1)\end{array},
\right.
\label{a3}
\end{eqnarray}
\begin{eqnarray}
&&\partial_t \bar{W}_2(x,y;\Lambda)
\nonumber\\
&=&-4\bar{W}_2(x,y;\Lambda)-2x\partial_x\bar{W}_2(x,y;\Lambda)
-2y\partial_y\bar{W}_2(x,y;\Lambda)
\nonumber\\
&&-\frac{1}{2\pi^2}f_Z^{-2}(1;\Lambda)\,
\bar{W}_2(x,1;\Lambda)\bar{W}_2(y,1;\Lambda)
+\frac{1}{48\pi^2}f_G^{-2}(1;\Lambda)\,
\bar{W}_1(x,1;\Lambda)\bar{W}_1(y,1;\Lambda)
\nonumber\\
&&+\frac{g^2(\Lambda_0)}{32\pi^2}f_G^{-2}(1;\Lambda)
f_G^{-1}({\rm max}(x,1);\Lambda)\,\bar{W}_1(y,1;\Lambda)\times
\left\{\begin{array}{ll}(3x^{-2}-2x^{-3})&(x>1)\\1&(x\le1)\end{array}
\right.
\nonumber\\
&&+\frac{g^2(\Lambda_0)}{32\pi^2}f_G^{-2}(1;\Lambda)
f_G^{-1}({\rm max}(y,1);\Lambda)\,\bar{W}_1(x,1;\Lambda)\times
\left\{\begin{array}{ll}(3y^{-2}-2y^{-3})&(y>1)\\1&(y\le1)\end{array}
\right.
\nonumber\\
&&+\frac{g^4(\Lambda_0)}{24\pi^2}f_G^{-2}(1;\Lambda)
f_G^{-1}({\rm max}(y,1);\Lambda)f_G^{-1}({\rm max}(x,1);\Lambda)
\nonumber\\
&&\quad\times\left(\frac{(5+3y)}{(1+y)^3}\right)\times
\left\{\begin{array}{ll}(3x^{-2}-2x^{-3})&(x>1)\\1&(x\le1)\end{array},
\right.
\label{a4}
\end{eqnarray}
\begin{eqnarray}
&&\partial_t \bar{W}_3(x,y;\Lambda)
\nonumber\\
&=&-4\bar{W}_3(x,y;\Lambda)-2x\partial_x\bar{W}_3(x,y;\Lambda)
-2y\partial_y\bar{W}_3(x,y;\Lambda)
\nonumber\\
&&-\frac{1}{48\pi^2}f_Z^{-2}(1;\Lambda)\,
\bar{W}_4(x,1;\Lambda)\bar{W}_4(y,1;\Lambda)+
\frac{1}{2\pi^2}f_G^{-2}(1;\Lambda)\,
\bar{W}_3(x,1;\Lambda)\bar{W}_3(y,1;\Lambda)
\nonumber\\
&&+\frac{9g^2(\Lambda_0)}{32\pi^2}f_G^{-2}(1;\Lambda)
f_Z^{-1}({\rm max}(x,1);\Lambda)\,\bar{W}_3(y,1;\Lambda)\times
\left\{\begin{array}{ll}x^{-2}&(x>1)\\1&(x\le1)\end{array}
\right.
\nonumber\\
&&+\frac{9g^2(\Lambda_0)}{32\pi^2}f_G^{-2}(1;\Lambda)
f_Z^{-1}({\rm max}(y,1);\Lambda)\,\bar{W}_3(x,1;\Lambda)\times
\left\{\begin{array}{ll}y^{-2}&(y>1)\\1&(y\le1)\end{array}
\right.
\nonumber\\
&&-\frac{g^2(\Lambda_0)}{64\pi^2}f_Z^{-2}(1;\Lambda)
f_G^{-1}({\rm max}(x,1);\Lambda)\,\bar{W}_4(y,1;\Lambda)\times
\left\{\begin{array}{ll}(3x^{-1}-x^{-2})&(x>1)\\(3-x)&(x\le1)\end{array}
\right.
\nonumber\\
&&-\frac{g^2(\Lambda_0)}{64\pi^2}f_Z^{-2}(1;\Lambda)
f_G^{-1}({\rm max}(x,1);\Lambda)\,\bar{W}_4(y,1;\Lambda)\times
\left\{\begin{array}{ll}(3y^{-1}-y^{-2})&(y>1)\\(3-y)&(y\le1)\end{array}
\right.
\nonumber\\
&&+\frac{9g^4(\Lambda_0)}{64\pi^2}f_G^{-2}(1;\Lambda)
f_Z^{-1}({\rm max}(x,1);\Lambda)f_Z^{-1}({\rm max}(y,1);\Lambda)
\nonumber\\
&&\quad\times\left(\frac{(1+y(6+y))}{(1+y)^4}\right)\times
\left\{\begin{array}{ll}x^{-2}&(x>1)\\1&(x\le1)\end{array}
\right.
\nonumber\\
&&-\frac{g^4(\Lambda_0)}{16\pi^2}f_Z^{-2}(1;\Lambda)
f_G^{-1}({\rm max}(x,1);\Lambda)f_G^{-1}({\rm max}(y,1);\Lambda)
\nonumber\\
&&\quad\times\left(\frac{1}{(1+y)}\right)\times
\left\{\begin{array}{ll}(3x^{-1}-x^{-2})&(x>1)\\(3-x)&(x\le1)\end{array},
\right.
\label{a5}
\end{eqnarray}
\begin{eqnarray}
&&\partial_t \bar{W}_4(x,y;\Lambda)
\nonumber\\
&=&-4\bar{W}_4(x,y;\Lambda)-2x\partial_x\bar{W}_4(x,y;\Lambda)
-2y\partial_y\bar{W}_4(x,y;\Lambda)
\nonumber\\
&&-\frac{1}{2\pi^2}f_Z^{-2}(1;\Lambda)\,
\bar{W}_4(x,1;\Lambda)\bar{W}_2(y,1;\Lambda)
+\frac{1}{2\pi^2}f_G^{-2}(1;\Lambda)\,
\bar{W}_3(x,1;\Lambda)\bar{W}_1(y,1;\Lambda)
\nonumber\\
&&-\frac{3g^2(\Lambda_0)}{8\pi^2}f_Z^{-2}(1;\Lambda)
f_G^{-1}({\rm max}(x,1);\Lambda)\,\bar{W}_2(y,1;\Lambda)\times
\left\{\begin{array}{ll}(3x^{-1}-x^{-2})&(x>1)\\(3-x)&(x\le1)\end{array}
\right.
\nonumber\\
&&+\frac{9g^2(\Lambda_0)}{32\pi^2}f_G^{-2}(1;\Lambda)
f_Z^{-1}({\rm max}(x,1);\Lambda)\,\bar{W}_1(y,1;\Lambda)\times
\left\{\begin{array}{ll}x^{-2}&(x>1)\\1&(x\le1)\end{array}
\right.
\nonumber\\
&&+\frac{3g^2(\Lambda_0)}{4\pi^2}f_G^{-2}(1;\Lambda)
f_G^{-1}({\rm max}(y,1);\Lambda)\,\bar{W}_3(x,1;\Lambda)\times
\left\{\begin{array}{ll}(3y^{-2}-2y^{-3})&(y>1)\\1&(y\le1)\end{array}
\right.
\nonumber\\
&&+\frac{3g^4(\Lambda_0)}{8\pi^2}f_G^{-2}(1;\Lambda)
f_G^{-1}({\rm max}(y,1);\Lambda)f_Z^{-1}({\rm max}(x,1);\Lambda)
\nonumber\\
&&\quad\times\left(\frac{(5+3y)}{(1+y)^3}\right)\times
\left\{\begin{array}{ll}x^{-2}&(x>1)\\1&(x\le1)\end{array}.
\right.
\label{a6}
\end{eqnarray}


\end{document}